\documentstyle[twocolumn,epsfig,prb,aps,amsmath,amssymb]{revtex}
\begin{document}

\twocolumn[\hsize\textwidth\columnwidth\hsize\csname
@twocolumnfalse\endcsname

\title{Interaction between impurities and solitons in quasi
one-dimensional spin-Peierls systems} 
\author{D.~Augier$^a$, P.~Hansen$^b$, D.~Poilblanc$^a$,
J.~Riera$^{a,b}$, and E.~S\o rensen$^a$}
\address{
$^a$Laboratoire de Physique Quantique \& Unit\'e Mixte
de Recherche CNRS 5626\\
Universit\'e Paul Sabatier, 31062 Toulouse, France\\
$^b$Instituto de F\'{\i}sica Rosario, Consejo Nacional de 
Investigaciones 
Cient\'{\i}ficas y T\'ecnicas y Departamento de F\'{\i}sica\\
Universidad Nacional de Rosario, Avenida Pellegrini 250, 
2000-Rosario, Argentina}
\date{\today}
\maketitle
\begin{abstract}
The role of the spin-phonon coupling in spin-Peierls chains
doped with spin-0 or spin-1 impurities is investigated 
by various numerical methods such as
exact diagonalization, quantum Monte Carlo simulations and
Density Matrix Renormalization Group.
Various treatments of the 
lattice, in a fully quantum mechanical way,
classically in the adiabatic approximation
or using a fixed three-dimensional dimerization pattern 
are compared. For an isolated chain, strong bonds form
between the two spin-$\frac{1}{2}$ sites
next to the impurity site, leading to the appearance of
magneto-elastic solitons. 
We also show that these excitations 
do not bind to spin-0 impurities but are weakly attracted by 
spin-1 impurities. However, the interchain elastic
coupling generates an effective confining potential at the 
non-magnetic impurity site which can lead to the formation of 
soliton-impurity bound states.
We also predict that a soliton and an antisoliton  bound to two
impurities on the same chain can annihilate each other
when the separation between the impurities is smaller than a 
critical value depending on the interchain elastic constant.


\end{abstract}

\vskip2pc]
%=====================================================================
% BODY OF PAPER
\section{Introduction}
\label{intro}

Quasi one-dimensional (1D) quantum antiferromagnets can exhibit 
surprising magnetic properties at low temperature. The recent
discovery
of the spin-Peierls (SP) transition in the inorganic compounds
CuGeO$_3$~(Ref.\onlinecite{hase}) and NaV$_2$O$_5$~
(Ref.\onlinecite{isobe}) has drawn both experimental and
theoretical interest. The chemistry  of these compounds 
indeed allows for the synthesis of large single crystals and 
consequently the achievement of new experimental studies~\cite{bore} 
which were not accessible to the previously known organic SP 
materials.
Nevertheless, these two compounds have quite different behaviors
presumably due to the fact that while NaV$_2$O$_5$ 
is believed to have quarter-filled ladders, 
CuGeO$_3$ is well described by weakly coupled 
spin-$\frac{1}{2}$ Heisenberg chains. In this sense, the present
studies are more related to this latter compound.

At a critical temperature, $T_c$, the SP compounds undergo
a phase transition driven by the spin-phonon coupling.
This SP transition, which is characterized 
by the opening of a spin gap and a lattice dimerization,
is experimentally
signaled by an isotropic drop of the magnetic susceptibility 
revealing the non-magnetic nature of the ground state (GS).
A general physical picture can be drawn from the consideration of
the exact GS of the frustrated Heisenberg chain 
at the so-called Majumdar-Ghosh point.
The spontaneously dimerized non-magnetic GS is
two-fold degenerate, corresponding to two possible dimerization
patterns A and B, which are a succession of disconnected 
singlet dimers. The elementary excitation called soliton
(antisoliton) consists of an unpaired spin separating A and B 
(B and A) patterns~\cite{shastry}. Solitons and antisolitons 
propagate, then acquiring a dispersion.

For temperatures above the critical temperature $T_c$,
quasi-1D SP compounds are usually described
as independent uniform AF Heisenberg chains, in some cases
including terms describing the frustration present in the system.
The nearest neighbor $J$ and next-nearest neighbor
$\alpha J$ exchange integrals can then be estimated by a fit of the
magnetic susceptibility~\cite{rieradobry,castilla}. The high
temperature behavior is governed by $J$ and the position of the
maximum by the frustration ratio $\alpha$. Values such as
$J\approx 160$~K and $\alpha \approx 0.36$ have been proposed for
CuGeO$_3$~(Ref.\onlinecite{rieradobry}).

CuGeO$_3$ can be easily 
doped by substituting magnetic Ni$^{2+}$ or
non-magnetic Zn$^{2+}$ impurities to spin-$\frac{1}{2}$ Cu$^{2+}$. 
A rapid suppression  of the spin gap due to impurity doping 
has been measured by magnetic susceptibility~\cite{hase2} 
and inelastic neutron scattering~\cite{lussier} experiments.
Competition between the antiferromagnetic (AF) and SP phases has also
been observed~\cite{hase2,oseroff,matsuda,koide,renard,regnault}.
Theoretical work analyzing the effects due to the doping by
vacancies has been carried out~\cite{martins,Laukamp,Hansen},
assuming a static dimerization.

The coupling to the lattice plays a major role in SP compounds. 
A description of these systems in terms of a static dimerized
Heisenberg chain is widely found in the literature. This approach
has some drawbacks: the dimerization is introduced {\it de facto}
in the model and is totally frozen, the lattice cannot adjust
to spin fluctuations, which are essential to understand the behavior
under high magnetic fields, or to inhomogeneities introduced by
impurities.

In this paper, SP spin-lattice models with spin-0 or spin-1 
impurities have been studied at $T=0$ or very low temperature 
using various numerical techniques such as Exact Diagonalization
(ED)~\cite{didier}, Quantum Monte-Carlo (QMC) and
Density Matrix Renormalization Group (DMRG). 
Emphasis has been put on the appearance of magneto-elastic
excitations and their interplay with impurities.
A strictly 1D dynamical spin-lattice model is
first studied in Sec.~\ref{1d}.
Non-uniform dynamical or adiabatic lattice distortions are 
considered and 
the existence of bound states due to impurities is investigated. 
The role of the three-dimensional character of the phonons 
is then considered, first in the simplest description in terms of
a static dimerized Heisenberg chain. 
Bound states between a soliton and a non-magnetic impurity  
are quantitatively studied in Sec.~\ref{dimerized}. A simple 
physical picture is also given to explain the spectrum of bound states. 
In Sec.~\ref{inter_cou} a more involved model is introduced by
considering explicitly the elastic coupling between the 1D
spin-phonon chains.
This model enables us to understand the physical origin of the
soliton-impurity bound states. 
%and allows us to extract some conclusions relevant for the 
%interpretation of some experimental results.
 
\section{1D spin-phonon chain}
\label{1d}

\subsection{Models}
The key ingredient of the physics of spin-Peierls compounds is
the dynamical spin-phonon coupling. 
A spin-$\frac{1}{2}$ excitation is expected to locally
distort the lattice, creating an elastic soliton.
Our starting point is the frustrated Heisenberg chain with the
parameters $J$ and
$\alpha$ defined below. In addition, we assume
a linear dependence of the nearest neighbor exchange integral
on the relative atomic displacement. Dispersionless 
phonons of frequency $\Omega$ 
are considered for sake of simplicity. One then 
obtains~\cite{Khomskii,Augier,affleck,hansen},

\begin{eqnarray}
{\cal H_\parallel^{\mathrm{(d)}}}=
   &J&\sum_{i} (1+g(b^{\phantom\dagger}_i+b_i^\dagger))
   (\vec{S}_{i} \cdot \vec{S}_{i+1}-\frac{1}{4}) \nonumber \\
&+&\alpha J \sum_j \vec{S}_{j} \cdot \vec{S}_{j+2}
+\Omega \sum_i \, b_i^\dagger b^{\phantom\dagger}_i,
\label{hamdy} 
\end{eqnarray}
\noindent
where $b_i^\dagger$ ($b_i^{\phantom\dagger}$) is the phononic
creation (destruction) operator at site $i$ and $g$ is the 
magneto-elastic coupling. Note that we have assumed that the
magneto-elastic coupling associated with the next-nearest
neighbor bonds can be neglected. 

The adiabatic limit of Hamiltonian~(\ref{hamdy}) is of
great interest.
When the phonon frequency
$\Omega$ is sufficiently small compared to
the spin fluctuation energy scale $J$, the
phonon degrees of freedom can be treated classically.
The Hamiltonian is then written as,

\begin{eqnarray}
{\cal H_\parallel^{\mathrm{(a)}}}=
J\sum_{i} (1&+&\delta_{i})\,
\vec{S}_{i} \cdot \vec{S}_{i+1}
+\alpha J \sum_j   \vec{S}_{j} \cdot \vec{S}_{j+2}\nonumber \\
&+& \frac{1}{2}K \sum_{i} \delta_{i}^2,
\label{hamad} 
\end{eqnarray}
\noindent
where $\delta_i$ is the distortion at site $i$, $K$ the string 
constant ($=\Omega/(2g^2)$) and the last term corresponds 
to the elastic energy loss. $\delta_i$ is expressed in units
so that the magneto-elastic coupling, which can be absorbed
in the definition of $K$, is set to unity.
It is interesting to notice that the frustration $\alpha$ alone
({\it i.e.} $g=0$) can lead to a dimerized state for
$\alpha > \alpha_c \simeq 0.241167$~
(Ref.~\onlinecite{haldane,crit,castilla,white,eggert}). 
In fact, the relevance of $\alpha$ is two 
fold; (i) it is required to explain qualitatively the properties of
the real compounds ($\alpha \simeq 0.36$ has been proposed in
CuGeO$_3$) and (ii) it can be used in numerical simulations to
reduce the spin-spin correlation length (for $\alpha=0.5$ and $g=0$,
it is only one lattice unit) and hence to reduce 
finite size effects.
 
\begin{figure}
\begin{center}
\epsfig{file=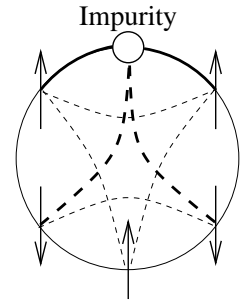,width=3cm}
\end{center}
%\vspace{0.2cm}
\caption{Example of a periodic chain with L=5 spin-$\frac{1}{2}$ 
sites and an additional impurity. Nearest 
neighbor (solid lines) and next nearest neighbor
(dashed lines) links are plotted. Bold links
correspond to the links between the impurity and its
spin-$\frac{1}{2}$ neighbors.}
\label{expl}
\end{figure}

The substitution of a spin-$\frac{1}{2}$ by 
a spin-S (S=0 or 1) impurity is then studied
using these Hamiltonians on chains with periodic boundary
conditions. The impurity is coupled to
its first and second nearest neighbors (see Fig.~\ref{expl}).
For sake of simplicity,
the couplings between two spins-$\frac{1}{2}$ or between 
a spin-S impurity and one of its spin-$\frac{1}{2}$
neighbors are taken identical, {\it i.e.} the 
same values of $J$, $\alpha$, $g$ and $\Omega$ are considered
all along the chain independently of the nature of the spins.
For instance, in the case of a spin-1 impurity, the Hamiltonian 
impurity part is explicitly

\begin{eqnarray}
{\cal H_{\mathrm{imp}}^{\mathrm{(d)}}}=
   &J& \sum_{l} (1+g(b^{\phantom\dagger}_l+b_l^\dagger))
   (\vec{S}_{l} \cdot \vec{s}-\frac{1}{2}) \nonumber\\
&+&\alpha J \sum_m \vec{S}_{m} \cdot \vec{s}
+ \Omega \sum_l \, b_l^\dagger b^{\phantom\dagger}_l
\label{eqimp}
\end{eqnarray}
\noindent
where the impurity spin operator is $\vec{s}$ and $l$ ($m$)
refers to its two nearest (next-nearest) neighbors. 
However, for a spin-0 impurity, the situation is simpler:
all links between the impurity and its spin-$\frac{1}{2}$ 
neighbors are taken as zero.
In this case, when $\alpha=0$, the system corresponds to
a chain with open boundary conditions (OBC).

\subsection{Dynamical model}
\begin{figure}
\begin{center}
\epsfig{file=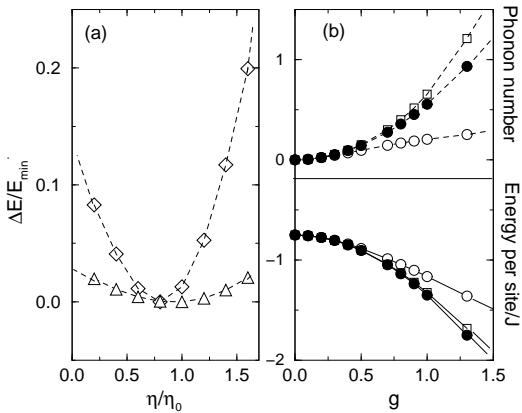,width=7cm}
\end{center}
\vspace{-0.1cm}
\caption{(a) Relative GS energy 
%$(E(\eta)-E_{\mathrm{min}})/E_{\mathrm{min}}$
of a 4-site chain as a function
of the variational parameter $\eta$ for $\alpha=0.5$,
$g=0.5$, $\Omega=J$ ($\triangle$) and $\alpha=0$, $g=1.$,
$\Omega=J$ ($\lozenge$). (b) GS energy (full line) and mean
phonon number (dashed line) per site for a 4-site
chain using a two ($\circ$), eight ($\bullet$) phonon states
truncation, and using the coherent state approximation with
$\eta=\eta_0$ ($\square$) as a function
of the magneto-elastic coupling $g$ for $\alpha=0$ and $\Omega=J$.}
\label{pho}
\end{figure}

{\it Numerical technique}:
A reliable numerical treatment of 
Hamiltonian~(\ref{hamdy}) is  
a difficult task. Indeed, the phononic Hilbert space is infinite,
even for a finite chain. Possible approaches such as QMC
simulations~\cite{sandvik}, single-mode 
approximation~\cite{augier}, or ED with a fixed maximum number 
of phonons~\cite{fewe} have been
proposed. The results obtained in this paper 
are based on an ED calculation using 
the coherent states introduced by Fehrenbacher~\cite{Fehrenbacher}.
Preliminary results have been reported
in Ref.~\onlinecite{Augier} in the case of pure systems
and Ref.~\onlinecite{hansen} for rings with impurities in
the absence of frustration. On each site,
a two state basis including the phononic vacuum $|0\rangle_i$
and a coherent state $|1\rangle_i$ is considered. The phononic
part of the Hamiltonian (\ref{hamdy}) can be rewritten as 
${\cal H}_{\mathrm{ph}}= J\lambda  \sum_i
(b^{\phantom\dagger}_i+b_i^\dagger) A_i+\Omega
\sum_i \, b_i^\dagger b^{\phantom\dagger}_i
$, where $\lambda$ is a
constant and $A_i$ an operator of eigenvalues 0 and 1
(the constant terms subtracted to $\vec{S}_i\cdot\vec{S}_j$ in
(\ref{hamdy}) and (\ref{eqimp}) have been introduced with this
purpose). It is convenient to introduce the coherent state 
$$|1\rangle_i=\exp{(-\eta^2/2)}
\exp{(\eta \ b_i^\dagger)}|0\rangle_i,$$
where $\eta$ is a variational parameter. In fact,
$|1\rangle_i$ is an eigenstate of the phononic Hamiltonian
for fixed $A_i=1$ provided that $\eta=\eta_0=-\lambda J/ \Omega$. 
It is shown in Fig.~\ref{pho}(a) that
$\eta_0$ is the best value of $\eta$ at small coupling.
Furthermore it is a reasonable choice 
for arbitrary coupling. Therefore we shall
assume $\eta=\eta_0$ in the remainder of the paper.
Note that the method could be improved by choosing explicitly
the optimum $\eta$ that minimizes the energy for each set
of parameters.
The comparison between this coherent state approach (using $\eta_0$)
and a truncation of the phononic Hilbert space retaining two or
eight phonon states at each site is
illustrated in Fig.~\ref{pho}(b)
for the GS energy and the mean phonon number 
per site. The results are much more accurate using coherent states
than when only keeping two phonon states even though

\noindent
both Hilbert spaces have the same sizes. Note that
although the GS energy is accurately obtained,
the coherent state basis seems 
to slightly overestimate the mean phonon number. However, even 
at large couplings, the phonon dynamics is qualitatively
preserved.  Consequently our non-perturbative
approach takes properly into account the effects of the phonons
and the results can be checked by comparing to
a full phonon calculation on small clusters (typically 
with $L=4$ sites) as in Fig.~\ref{pho}(b).
This variational treatment of the dynamical phonons hence enables us to 
handle clusters with up to $L=16$ sites.
Special care is needed for the treatment of the bonds connected to the
spin-1 impurity. Indeed, special variational parameters 
(still corresponding to $\eta_0$) have
to be chosen for these bonds because the corresponding $A_i$ operators 
for the phononic part of the Hamiltonian differ from those of
the rest of the chain.

{\it Results}:
Let us summarize the main properties of this model. It has been 
shown that the GS of this model undergoes
at $T=0$ a spontaneous symmetry breaking toward a dimerized
gapped phase in a large region of parameter space~\cite{Augier}.
The elementary excitations are characterized as topological solitons,
{\it i.e.} unpaired spins separating the two different
dimer patterns.
Furthermore solitons ($s$) and antisolitons ($\bar{s}$) do 
not bind. More explicitly, one can
define the $s\bar{s}$ binding energy as the $p \rightarrow \infty$
extrapolation of

\begin{eqnarray}
E_B^{s\bar s}(L=2p)=[ E_0(2p,1)-E_0(2p,0)]
-2e_s,
\label{bind}
\end{eqnarray}
\noindent
where $E_{0}(l,S_z)$ is the GS energy of a $l$-site periodic chain
in the $S^z$ sector. If $l$ is even and $S^z=0$, $E_{0}(l,S^z)$ 
corresponds in fact to the ``vacuum'' energy, the energy of the pure
chain without any topological defect. 
$e_s$ is the soliton (antisoliton) minimum energy, defined as 
$e_s=\lim_{p\rightarrow \infty} [E_0(2p+1,\frac{1}{2})-E_0^*(2p+1)],$
where $E_0^*(2p+1)$ is the interpolation between 
$E_0(2p,0)$ and $E_0(2p+2,0)$. Thus, $e_s$ results as the energy
difference between an odd chain containing a soliton
and an (hypothetical) odd chain 
without soliton. Similarly, $E_B(s\bar s,L)$ is the energy 
difference between a chain with a $s\bar s$ pair and
two isolated solitons.
Previous studies provide strong
evidences that the $s\bar s$ binding energy vanishes~\cite{Augier},
{\it i.e.} $E_B=0$ in the thermodynamic limit in the absence of any
explicit dimerization.
Similar forms of binding energy will be discussed in the
context of impurity-soliton binding.
 
\begin{figure}
\begin{center}
\epsfig{file=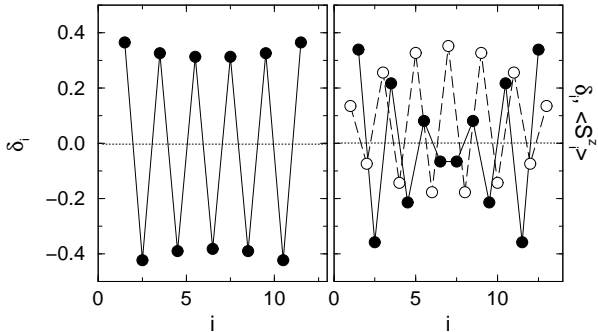,width=8cm}
\end{center}
%\vspace{0.2cm}
\caption{Distortion $\delta_i$ ($\bullet$) and z-component of the
spin $\langle S^z_i \rangle$ ($\circ$) as
a function of the site i on (a) a 12-site and (b) a 13-site
chain for $\alpha=0.36$, $g=0.8$ and $\Omega=J$ with OBC
using ED.}
\label{pattdyn}
\end{figure}

We first consider the case of a non-magnetic impurity.
Typical patterns are shown in Fig.~\ref{pattdyn}(a) for parameters
corresponding to CuGeO$_3$.
For even chains, the distortion pattern $\delta_i=
g\langle b_i^\dagger+b_i^{\phantom\dagger} \rangle$ rapidly oscillates
with a two site period. The equilibrium pattern
corresponds to strong bonds ({\it i.e.} $\delta_i$ positive)
next to the impurity. The amplitude $|\delta_i|$ varies
along the chain with increasing magnitude for decreasing
distance to the impurity. In other words, the impurity 
being located on site 0, the strongest bonds are the ones connecting
sites 1 and 2 and sites L-1 and L. In addition,
the average magnitude of the modulation increases with a stronger 
magneto-elastic coupling or weaker phonon frequency.
The z-component of the spin $\langle S^z_i \rangle$ is zero because
the chain is in a dimerized singlet state.
For odd chains (Fig.~\ref{pattdyn}(b)), the presence of 
a soliton at the center of the chain
is signaled by a point where $\delta_i$ vanishes and 
$\langle S^z_i \rangle$ is maximum. This corresponds to the 
characteristic schematic
picture of the soliton as an unpaired spin separating
two dimer patterns.
Similarly to the even chain case, strong bonds 
also form at the chain edges. The finite frustration that we have 
considered here does not seem to change the main features of the
pattern (see Ref.\onlinecite{hansen} for comparison). The fact
that the soliton forms at the center of the chain suggests that,
contrary to the conclusions of Ref.\onlinecite{Fukuyama}, solitons do
not bind to non-magnetic impurities in pure 1D spin-lattice models.

In order to be more quantitative,
we define a soliton-impurity binding energy 
similarly to (\ref{bind}),

\begin{eqnarray}
E_B^{I-s}(L=2p+1)&=&[E'_{0}(2p+1,\frac{1}{2})-E_0^*(2p+1)]\nonumber\\
&-&e_s-e_I,\label{bindi}
\end{eqnarray}
\noindent
where $E'_{0}(l,S^z)$ is the GS energy in
the $S^z$ sector of a chain with
$l$ spin-$\frac{1}{2}$ sites and with an extra impurity.
$e_I$ is an impurity minimum energy, 
$e_I=\lim_{p\rightarrow \infty} [E'_0(2p,0)-E_0(2p,0)],$ which
indeed corresponds to the energy difference between a chain with
and without an impurity. 

\begin{figure}
\begin{center}
\epsfig{file=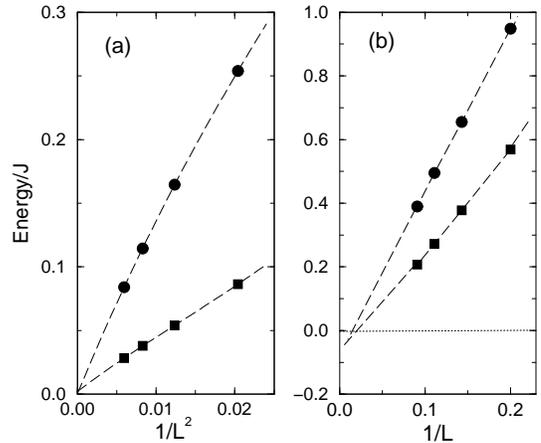,width=7cm}
\end{center}
%\vspace{0.2cm}
\caption{Soliton-impurity binding energy using ED for $\alpha=0.5$, 
$g=0.4$, $\Omega=0.3J$ ($\blacksquare$)
and $\alpha=0$, $g=0.45$, $\Omega=0.3J$ ($\bullet$) 
(a) as a function of the inverse square chain length $\frac{1}{L^2}$
in the case of a spin-0 impurity; (b) as a function of the inverse
chain length $\frac{1}{L}$ in the 
case of a spin-1 impurity.}
\label{sca_bind}
\end{figure}

The results shown in Fig.~\ref{sca_bind} (a)
reveal no binding for any finite chain.
Note that finite size effects are small
for parameters in the vicinity of the MG point because
of the small magnetic correlation length present in this case.

The same calculations can be performed in the case of a spin-1
impurity. The different scaling behaviors in Fig.~\ref{sca_bind}
($\frac{1}{L^2}$ for the spin-0 
and $\frac{1}{L}$ for the spin-1 impurity) 
suggest that the physics involved is also different.
Although a $\frac{1}{L^2}$ behavior may also occur
for larger rings with a spin-1
impurity, our data are nevertheless compatible 
with a small binding between the spin-1 impurity and the soliton.

\subsection{Adiabatic model}\label{adiab}

{\it Numerical technique}:
As seen above, a full treatment of the lattice dynamics is tedious
and restricted to small clusters. On the other hand,
when $\Omega \rightarrow 0$, an adiabatic approximation of 
the lattice is justified and enables us to handle larger clusters
both in ED or QMC calculations.
The main technical issue is to find the set of 
distortions $\{\delta_i\}$ which minimizes the total energy,
$\frac{\partial \langle {\cal H_\parallel^{\mathrm{(a)}}} \rangle}
{\partial \delta_i}=0,$ which leads to  
\begin{eqnarray}
J\langle \vec{S}_{i} \cdot \vec{S}_{i+1}\rangle
+K\delta_i=0,\label{eq} 
\end{eqnarray}
where $\langle\cdots\rangle$ is the mean value in the GS for an ED
calculation or some thermal average for a QMC calculation. 
Eq. (\ref{eq})
must be solved with the constraint of zero total distortion,
\begin{eqnarray}
\sum_i \delta_i=0 \label{constraint}.
\end{eqnarray}
This problem is clearly
self-consistent since in the equilibrium condition~(\ref{eq})
the distortion pattern is present in both terms: implicitly in the
first one because the GS depends on it, and explicitly
in the second one.
Consequently this problem can be treated
using the following iterative procedure.\cite{feiguin}
First, a randomly chosen set $\{\delta_i^0\}$ satisfying
(\ref{constraint}) is taken.
Applying the equilibrium condition~(\ref{eq}),
where the mean values are considered
using the adiabatic Hamiltonian with the set $\{\delta_i^0\}$, gives
another set $\{\delta_i'\}$ of distortions.
After subtracting its mean value,
a new set $\{\delta_i^1\}$ satisfying (\ref{constraint})
is then obtained and
the procedure is repeated until convergence. 

This scheme can be applied
using ED or QMC techniques.  Note however that in
this last case, although larger lattice sizes can be 
considered, one is restricted to the
non-frustrated $\alpha=0$ case to avoid the minus sign problem.
As expected, for a (pure) periodic chain, the static uniform
dimerization pattern $\delta_i=
(-1)^i \delta$ is obtained through this procedure.

{\it Results}:
In this part, we shall study  the introduction of a spin-0 or a
spin-1 impurity in a chain described by the adiabatic Hamiltonian.
It is here crucial to enable
the magnitude of the modulation $|\delta_i|$
to vary along the chain. A QMC self-consistent method is used so
that we shall restrict ourselves to 
the non-frustrated $\alpha=0$ case in order to avoid sign problems.
A standard
world-line algorithm was implemented. Most of the calculations were
performed at $T = 0.05 J$ which we have shown in previous studies is
low enough to reflect essentially the GS
properties.\cite{feiguin,Laukamp,Hansen} Besides, since we
are mostly interested in GS properties we have carried on the
simulations in the subspace of minimum $z$-component of the
total spin. The number of slices in the Trotter direction was taken
equal to $M=120$.

\begin{figure}
\begin{center}
\epsfig{file=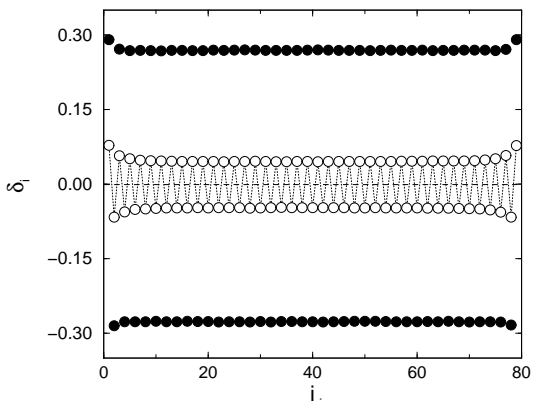,width=7cm}
\end{center}
\vspace{-0.2cm}
\caption{Distortion patterns $\delta_i$ on a 80 site chain as
a function of the position i
for $\alpha=0$, $K=J$ ($\bullet$, the dotted line between the
computed points
is not shown for clarity) and $K=3J$ ($\circ$) using
QMC ($T=0.05J$).}
\label{disteven}
\end{figure}

\begin{figure}
\begin{center}
\epsfig{file=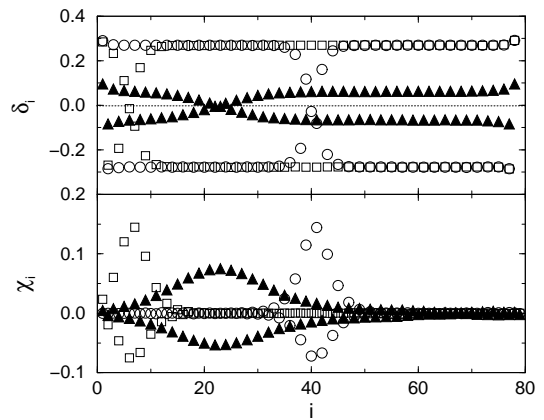,width=7cm}
\end{center}
\vspace{-0.2cm}
\caption{Distortion $\delta_i$ (a) and magnetic susceptibility 
$\chi_i$ (b) patterns as a function of the position i
on a 79 site chain for 
$\alpha=0$, $K=J$ ($\circ$, $\square$ correspond to two different
runs) and $K=2.5J$ ($\blacktriangle$) using
QMC ($T=0.05J$). Lines joining computed points are not
plotted for clarity.}
\label{distodd}
\end{figure}

Patterns of dimerization are shown in 
Figs.~\ref{disteven} and \ref{distodd} for
a chain with a spin-0 impurity and are similar
to those observed using the dynamical approach. 
For even chains, the main features are
identical (Fig.~\ref{disteven}). 
The distortion rapidly oscillates with a two-site period 
with strong bonds on the edges, where the impurity lies. Its
amplitude is slightly varying along the chain, the 
strongest distortion magnitude being observed at the edges.
The magnitude of the distortion decreases when
$K$ increases as seen in Fig.~\ref{disteven}. The z-component
of the spin $\langle S^z_i \rangle $ vanishes in this case because 
the chain is in the singlet dimer SP state.

For odd chains, a single solitonic excitation is located away
from the edges as shown in Fig.~\ref{distodd}. Its presence 
is inferred, as previously, from a local vanishing $\delta_i$ and 
a maximal $\langle S^z_i \rangle$. 
In this case we consider the local spin susceptibility,
\begin{eqnarray}
\chi_i=\frac{1}{T} \sum_j \langle S^z_i S^z_j \rangle
\end{eqnarray}
instead of $\langle S^z_i
\rangle$. Indeed, when using a QMC algorithm
with an odd number of sites, a fictitious site has to be added
and the total spin of the chain can fluctuate 
between $+\frac{1}{2}$ and $-\frac{1}{2}$ through spin flips
with this additional spin.
However, the spin susceptibility $\chi_i$ deals with a z-component
of the local spin with respect to the chain {\it global} spin 
orientation~\cite{eggert,Laukamp}. Note that, 
however, this excitation is topological in nature in the sense 
that $\chi_i$ integrated in space on a finite region
gives exactly a Curie law of a spin-$\frac{1}{2}$.
The soliton width increases as the spin-phonon coupling decreases 
and becomes a sinusoidal distortion in the weak
coupling limit.\cite{feiguin} In fact, the inverse soliton width,
the spin gap, and the amplitude of the dimerization
are three different features of the SP transition and all of them
increase with the spin-phonon coupling.

It is interesting to
notice that different QMC runs ($\circ$ and $\square$ in
Fig.~\ref{distodd}) lead to the very same solitonic patterns
centered at different sites in a large area around the middle
of the chain. This is another evidence that there
is no binding between non-magnetic impurities and solitons
in this strictly one-dimensional model. The fact that the solitons
are at least a distance away of the edges equal to half the
soliton width indicates a likely short range repulsion between
the soliton and the impurity.

\begin{figure}
\begin{center}
\epsfig{file=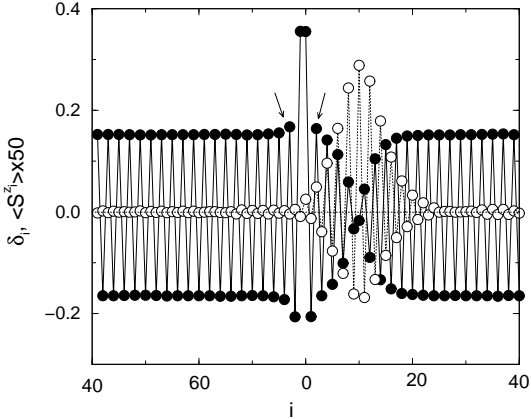,width=7cm}
\end{center}
\vspace{-0.2cm}
\caption{Local distortion $\delta_i$ ($\bullet$) and z-component of
the local spin $\langle S^z_i \rangle$ ($\circ$) as
a function of the position i
on a 79 site chain for $\alpha=0$ and $K=1.5J$
using QMC ($T=0.05J$). A spin-1 impurity
is located at site 0 at the middle of the figure 
and periodic boundary conditions are used. Two arrows indicate 
particularly strong
bonds ($\delta_i>0$) next to the 
spin-0 object formed by the spin-1 and its two
neighbors.}
\label{spin1}
\end{figure}

Next, we turn to the case of a spin-1 impurity.
Typical distortion and susceptibility patterns are shown in 
Fig.~\ref{spin1} for an odd length chain. In first approximation, the
spin-1 impurity and its two spin-$\frac{1}{2}$ neighbors behave
like a
singlet,\cite{eggert2} {\it i.e.} like a spin-0 impurity as
indicated by very strong bonds connected to the spin-1 impurity. 
This three-spin object is weakly connected to the rest of the system
as indicated by particularly weak bonds ($\delta_i<0$) between sites
1 and 2, and L-1 and L (the impurity is located at
site 0).\cite{Hansen} Furthermore,
strong bonds can even be observed close to this composite object
(see arrows in Fig.~\ref{spin1}) as it is the case
for a real S=0 impurity.
Physically, the three-spin 
cluster spends most of the time in a singlet state configuration.
One observes that the soliton
always stays quite close to the impurity, independently
of the QMC run, 
suggesting a small residual impurity-soliton binding,
consistent with the previous dynamical calculation.

\section{Explicitly dimerized chains}
\label{dimerized}

The Hamiltonians considered in Sec.~\ref{1d} are purely
one-dimensional. However, in SP compounds, 
the phonons have a three-dimensional character.   
One of the simplest models proposed in the
literature to take this
three-dimensional behavior into account
is the dimerized Heisenberg chain,
\begin{eqnarray}
{\cal H_{\mathrm{MF}}}=J\sum_i [(1+\delta(-1)^i)\vec{S}_i \cdot
\vec{S}_{i+1}+\alpha \vec{S}_i\cdot \vec{S}_{i+1}],\label{dimer}
\end{eqnarray} where $\delta$ is an explicit dimerization. 
Contrary to the models considered in Sec.~\ref{1d},
the distortion pattern is here static and uniform.
It is interesting to note that this 
Hamiltonian (\ref{dimer}) can in fact be inferred from 
the dynamical Hamiltonian (\ref{hamdy}) when only the 
dominant $\pi$-mode phonon is considered~\cite{augier}. 

\begin{figure}
\begin{center}
\epsfig{file=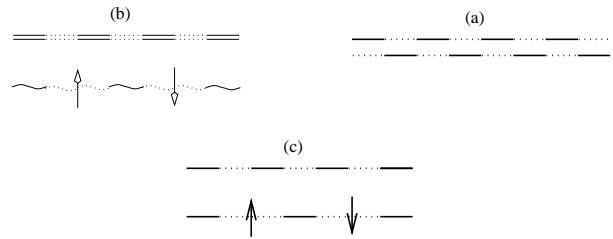,width=8cm}
\end{center}
\caption{(a) The two-fold degenerate GS of Hamiltonian
(\ref{hamdy}). The solid (dashed) lines indicate strong (weak) bonds
and dimer singlets. (b) The fixed dimerization pattern of 
${\cal H_{\mathrm{MF}}}$ (above) and a magnetic configuration of 
singlet dimers
with a $s\bar{s}$ pair (below). (c) A magneto-elastic state without
(above) and with a $s\bar{s}$ pair (below) of Hamiltonian
(\ref{hamad}).
}
\label{last}
\end{figure}

The dimerized Heisenberg chain has been widely studied in
the literature. Its main features
are the following: the elementary excitations can
be interpreted as magnetic solitons and antisolitons (or spinons), 
namely isolated spins-$\frac{1}{2}$ separating two dimerization
patterns~\cite{shastry,affleck}. As in the models considered in the
previous section, for even site chains, solitons and antisolitons
do not exist in this model as independent particles but appear in 
pairs. However, in this case, as a difference to those models, they
are confined by a linear potential proportional to 
$\delta (\ne 0)$ in the weak $\delta$
limit~\cite{affleck,patch,uhrig}. It has also been shown that the
excitation spectrum is constituted of a ladder of soliton-antisoliton
bound states lying below a two-magnon
continuum~\cite{affleck,sorensen,bouzerar}. The differences between
the models studied in the previous section and the present one can be
schematically visualized in the large dimerization limit as
in Fig. \ref{last}. While in the fully quantum dynamical model both
the distortion pattern $\{\delta_i \}$ and magnetic dimers are
two-fold degenerate, in the chain with fixed dimerization only one of
the distortion patterns is chosen and a magnetic state with $s\bar{s}$
pair separating two magnetic dimers configuration would leave a
region with a ``wrong" pattern in between the soliton and anti-soliton
thus leading to a confining potential.
On the other hand, this confining is absent if Hamiltonian
(\ref{hamad}) is considered since in this case the lattice would
relax following the magnetic order.

\begin{figure}
\begin{center}
\epsfig{file=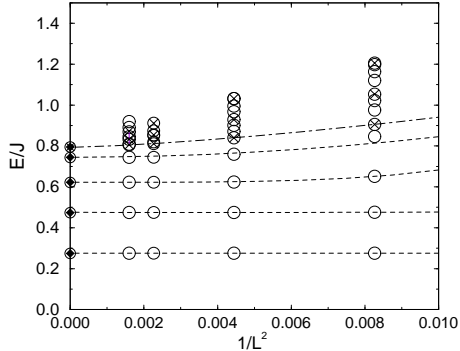,width=6cm}
\end{center}
\vspace{-0.2cm}
\caption{Lowest lying $S^z=\frac{1}{2}$ ($\circ$) and
$S^z=\frac{3}{2}$ (X) excitations for the $\alpha=0.5$ and
$\delta=0.05$ Heisenberg chain as a function of the square inverse
of the chain length L (up to L=25) obtained by ED. The energy
reference is the GS of the even periodic chain interpolated to
odd sizes. Dashed lines represent
a $(\frac{1}{L^2},\frac{1}{L^3})$ fit and the extrapolations
to infinite sizes are indicated. DMRG extrapolations
are also plotted ($\blacklozenge$). The dot-dashed line indicates
the onset of the continuum. }
\label{spectre}
\end{figure}

The case of an odd chain with open boundary conditions,
{\it i.e.} with a spin-0 impurity cutting the chain, has been
extensively studied in Ref.\onlinecite{Laukamp}. It has been shown
that, due to the linear confining potential, an 
isolated soliton binds to the spin-0 impurity next
to a weak bond~\cite{Laukamp,uhrig,sorensen}.
In this section, we
generalize the approach of Ref.~\onlinecite{Laukamp}
to investigate the possibility
of several bound states and to study the low energy spectrum.
Following Ref.~\onlinecite{sorensen}, we assume that the
dimerization is stabilized by a strong frustration $\alpha \sim 0.5$
while the ``confining force'' introduced by $\delta$ remains
small. The excitation spectrum with respect to the ``vacuum'' 
energy {\it i.e.} the GS energy of a periodic even chain (without
any defect) interpolated to odd chains, is shown
in Fig.~\ref{spectre}. The energies could also be equivalently
measured with respect to the beginning of the continuum
but finite size effects affect its position
making results more dependent of the lattice size. 
For $\alpha=0.5$ and $\delta=0.05$, four soliton-impurity bound
states below the continuum can be clearly 
identified. The energy of the soliton-impurity
bound state can increase up
to an upper bound, above which
it becomes energetically more favorable to
create a soliton-antisoliton bound state. This state, which is a 
combination of a soliton-impurity and soliton-antisoliton
bound states, has a spin $\frac{3}{2}$ and 
is the lowest state of the continuum. We have checked that its
energy $E_0(\frac{3}{2})=(0.7949\pm 10^{-4})J$, 
is in the thermodynamic limit the sum of the lowest
impurity-soliton bound state $E_0(\frac{1}{2})=(0.2748\pm 10^{-4})J$ 
plus the lowest energy of 
a $s\bar{s}$ bound state, {\it i.e.} the spin gap 
$\Delta^{01}=(0.5200\pm 10^{-4})J$, where
$\Delta^{01}= E_0(2p,1)-E_0(2p,0)$
in the notation of Eq. (\ref{bind}).
The above energies are infinite size extrapolations obtained
with ED and DMRG for $\alpha=0.5$ and $\delta=0.05$.

\begin{figure}
\begin{center}
\epsfig{file=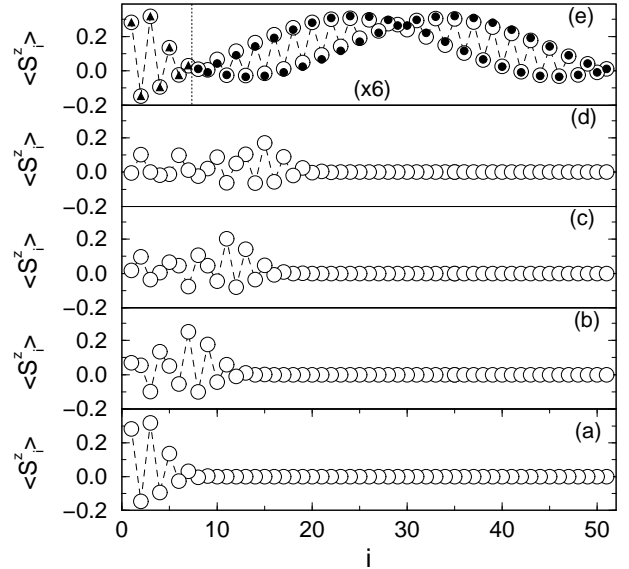,width=8cm}
\end{center}
\vspace{-0.2cm}
\caption{$\langle S^z_i \rangle$ as the function
of the site i for the four lowest spin-$\frac{1}{2}$ (a,b,c,d)
and the lowest spin-$\frac{3}{2}$ state (e) for
$\alpha=0.5$, $\delta=0.05$ on a 51-site chain using a DMRG
algorithm. The bond between the first two sites is ``weak".
In (e), a magnification by a factor 6 has been applied to the
points left to the dashed line. The 7 first sites
of the lowest bound state (a) ($\blacktriangle$) and a 
soliton-antisoliton pair on a 44 site lattice ($\bullet$) are also
shown for comparison.}
\label{etats}
\end{figure}

The wave functions $\langle S^z_i \rangle$ obtained in a DMRG
calculation shown in
Fig.~\ref{etats} clearly support this scenario. The four lowest states
(a,b,c,d) show a soliton bound to the weak bond edge, and 
the soliton moves further away from the impurity
when its energy increases. In contrast, in the lowest 
spin-$\frac{3}{2}$ state (Fig.~\ref{etats}(e)), 
a $s\bar{s}$ bound pair can be clearly identified in addition to
the solitonic GS seen in Fig.~\ref{etats}(a). 
The number of soliton-impurity bound states below the continuum
seems to increase with $\frac{1}{\delta}$, 
as was originally 
proposed for the number of $s\bar{s}$ pairs~\cite{affleck,sorensen}.

\section{Elastic interchain coupling}
\label{inter_cou}

The three-dimensional character of phonons
has been considered in Sec.~\ref{dimerized} using
the dimerized Heisenberg chain.
However this can be achieved in a more appropriate way
by adding to the pure 1D models of Sec.~\ref{1d} an
interchain coupling. Such an approach is particularly
necessary for materials with large anisotropy in the
elastic properties. Our motivation here is to improve the description 
of Sec.~\ref{dimerized} in order to enable the lattice to locally 
relax following the magnetic order. As seen previously in the 
case of isolated chains, such effects are crucial 
in the presence of inhomogeneities introduced by local magnetic
excitations (solitons) or impurities and lead to  
important qualitative differences with the dimerized Heisenberg chain 
(\ref{dimer}).

In a quasi-1D SP compound,
a given chain cut at its ends by spin-0 impurities (``impurity chain")
is immersed in the bulk. The neighboring chains are in the SP phase,
i.e. they have a uniform dimerization,
and produce on this chain a $q=\pi$ potential through an
elastic interchain coupling of the form~\cite{Khomskii,affleck},
$${\cal{H}_\perp}=K_\perp \sum_i \delta_i \delta_i^{\mathrm{ext}},$$
where $\delta_i$ is the distortion of the impurity chain,
$\{\delta_i^{\mathrm{ext}}\}$ are associated to the neighboring
chains and $K_\perp$ is the interchain elastic coupling constant.
Note that using this model, two neighboring chains 
can be in phase or out of phase
depending on the sign of $K_\perp$.
The neighboring chains can be treated in the mean-field approximation
while fully retaining the dynamics of the considered chain:
$\delta_i^{\mathrm{ext}}=(-1)^{i+1} \delta_0$. Consequently, one 
obtains,

$${\cal{H}_\perp}= K' \sum_i (-1)^{i+1} \delta_i,$$

\noindent
where $K'=K_\perp \delta_0$.

\begin{figure}
\begin{center}
\epsfig{file=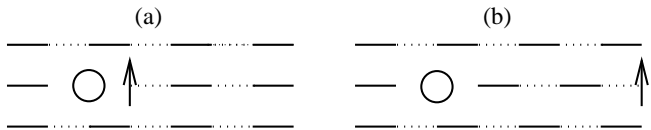,width=8.5cm}
\end{center}
\caption{Introduction of a non-magnetic impurity
in a three-dimensional spin-Peierls compound.
Schematic pattern pictures are drawn for the case of $K' > 0$.
(a) corresponds to a strong external potential and (b) to a
weak external potential. Full (dashed) lines symbolize strong (weak)
bonds.}
\label{patt}
\end{figure}

As schematically shown in Fig.~\ref{patt}(a), in a strong external
potential, the situation is similar to the one of Sec.~\ref{dimerized}:
the soliton is bound to the spin-0 impurity at the weak bond edge.
However, in a weak external potential, results of Sec.~\ref{1d}
suggest that solitons can delocalize away from the spin-0
impurity edge as depicted in Fig.~\ref{patt}(b) leaving strong bonds
at the edges. Our aim here is to study how, when switching on an
external potential and increasing it, the soliton 
binds to the impurity. The results shown in this part
are obtained using an adiabatic treatment of the lattice
and a QMC simulation of the spin degrees
of freedom (similar to Sec.~\ref{adiab}) for non-frustrated chains
with spin-0 impurities. Consequently, the lattice
can be seen as open chains with two unequivalent non-magnetic
impurities at its edges.

We first consider the case of an odd size chain. For vanishing 
$K'$, the soliton is free and strong bonds
($\delta_i > 0$) occur at the two chain ends.  
A non-zero $K'$ creates a linear
confining potential between the soliton and the impurity where
the potential tends to form a weak bond. This confining
potential then equilibrates with the short range impurity
repulsion. Consequently, the soliton moves to-

\begin{figure}
\begin{center}
\epsfig{file=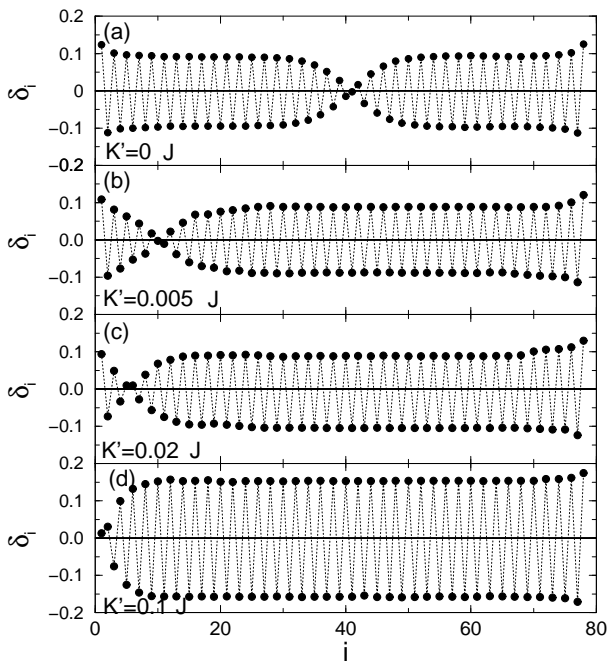,width=8cm}
\end{center}
\vspace{0.3cm}
\caption{Lattice distortions $\delta_i$ as
a function of the position i on a 79 site chain for $\alpha=0$,
$K=2J$ with an interchain coupling $K'$ obtained by QMC simulation 
($T=0.05J$). The value of $K'$ is indicated on the graph.}
\label{potexto}
\end{figure}

\noindent
wards the impurity
when the interchain coupling $K'$ increases, as can be seen in
Fig.~\ref{potexto}. Moreover, the magnitude of the 
dimerization increases with $K'$ as expected.
Note that the distortion pattern obtained in Fig.~\ref{potexto}(d),
{\it i.e.} stronger bond on one edge and zero distortion at the
other edge, 
is strikingly different from those of Ref.\onlinecite{Fukuyama}.
These two different behaviors (obtained for $i\sim 1$ and 
$i \sim L$) are in fact expected on each side of any given
impurity in this system.

\begin{figure}
\begin{center}
\epsfig{file=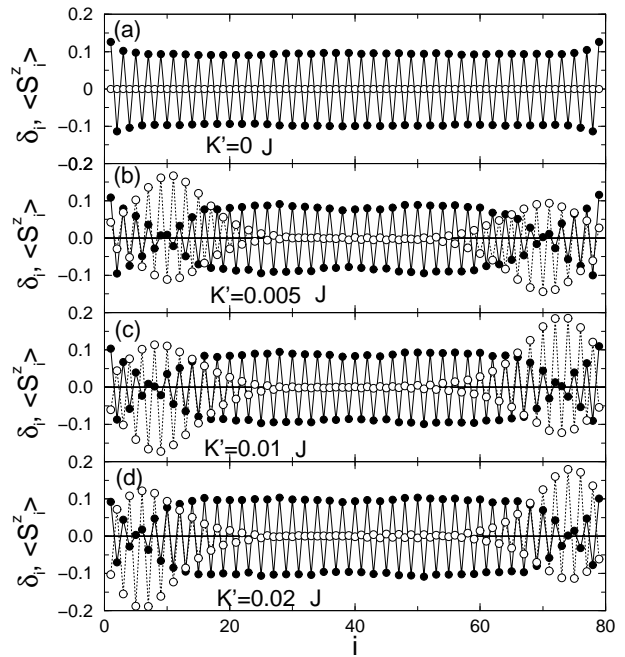,width=8cm}
\end{center}
\vspace{0.2cm}
\caption{Lattice distortions $\delta_i$ ($\bullet$)
and z-component of spin $\langle S^z_i\rangle$ ($\circ$) as
a function of the position i
on a 80 site chain for $\alpha=0$,
$K=2J$ with an interchain coupling $K'$ using
QMC ($T=0.05J$). The value of $K'$ is
indicated on the graph.}
\label{potexte}
\end{figure}

For even size lattices, two cases have to be examined.
If the free distortion pattern and the external potential
are in phase ({\it i.e.} if $K' \delta^0_i
\delta_i^{\mathrm{ext}}>0$, where $\delta^0_i = \delta_i(K'=0)$ ),
the external potential amplifies 
the free distortions seen in Fig.~\ref{disteven}
with strong bonds on the edges.
However, if they are out of phase, the external potential
will create two topological soliton defects. These
excitations move to the chain edges and become more localized
when the external potential increases, similarly to
what happens for a static dimerized chain~\cite{Laukamp,sorensen}.
The patterns shown in Fig.~\ref{potexte} support this scenario:  
a spin $\pm \frac{1}{2}$ is located near each edge.
Note that at low temperature, this excess of spin-$\frac{1}{2}$ at
the chain's ends oscillates between positive and negative value
for different QMC runs, corresponding to the
$S^z \rightarrow -S^z$ symmetry.
Consequently the role of the interchain elastic coupling 
is crucial in order to stabilize impurity-soliton bound states.

The formation of the $s\bar s$ topological defect in the open
chains with an even number of sites $L$ (as described above) is 
particularly subtle. In fact, because of the finiteness of
the energy cost $\Delta^{01}$ (or $2e_s$ if the solitons are
far away) associated with the formation of the defect, one can,
on general grounds, deduce the existence of a critical value
$K'_c$ of $K'$ in such a phenomenon. A comparison between the
energy cost $\propto 2e_s \simeq \Delta^{01}$ and the 
transversal elastic energy gain $\propto K'(L- 2 \Gamma)$, 
where $\Gamma$ is half the soliton width, leads to 
$K'_c \sim \Delta^{01}/(L- 2 \Gamma)$. Since $\Delta^{01}$
increases with decreasing $L$, and $ \Gamma =  \Gamma(K)$ is
roughly independent of $L$ (see Ref. \onlinecite{dobryriera}), 
then we expect that $K'_c $ decreases
for increasing chain length. For $K = 2$ we have obtained
$K'_c = 0.050, 0.032 \rm ~and~ 0.005$ for 
$L = 20, 40 \rm ~and~ 80$
respectively, thus confirming this prediction.

For real SP compounds doped with non-magnetic impurities, 
chains are cut at random places. 
By simple inspection of all possible
configurations, the previous study shows unambiguously that
exactly one spin-$\frac{1}{2}$ becomes bound to each impurity due
to the interchain elastic coupling responsible for 
the three-dimensional character of the lattice dynamics
provided that $K' > K'_c$.
However, in chains short enough that $K'_c(L) > K'$, 
non-magnetic impurities at the ends do not carry magnetic moments.
In other words, let us assume that one, by thought, moves
closer to each other two impurities with respectively a 
soliton and anti-soliton bound to them.
Our previous analysis shows that, below a critical separation
$L_c\simeq 2\Gamma + \Delta^{01}/K^\prime$ between the two impurities, 
this magnetic state becomes unstable towards a new
configuration where the soliton and the antisoliton annihilate each 
other and leave behind a dimerization pattern in antiphase with the 
3D order. Notice that such open chains are 
most likely to appear in the system after doping since 
the probability distribution of lengths is $P(L) = c (1-c)^L$, 
where $c$ is the impurity doping.
Furthermore, the characteristic distance $L_c$ can become fairly large
for small interchain elastic coupling $K^\prime$.

\begin{figure}
\begin{center}
\epsfig{file=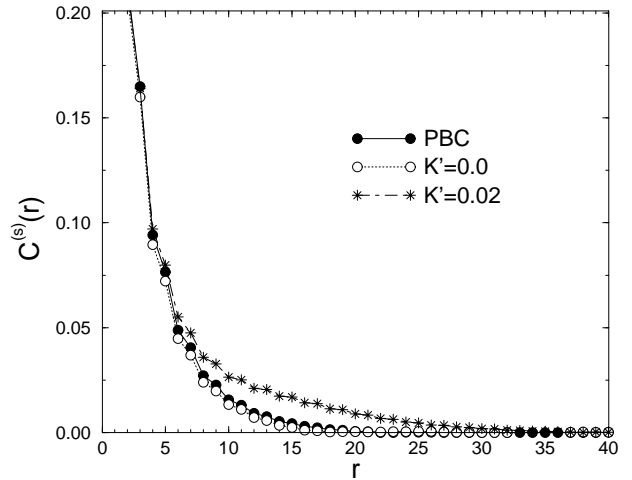,width=8cm}
\end{center}
\vspace{0.2cm}
\caption{Staggered spin-spin correlations in the 80 site chain
with $K=2$ obtained by QMC ($T=0.050$). 
Periodic boundary conditions with $K'=0$ and OBC with 
various values of the interchain coupling $K'$ as
indicated on the plot have been considered.
}
\label{cor80kp}
\end{figure}

Finally, in Fig. \ref{cor80kp} we show the staggered spin-spin 
correlation function $C^{(s)}(|i-j|)=(1/L)\sum_{i,j}\big<\vec{S}_i\cdot
\vec{S}_j\big>(-1)^{|i-j|}$ (normalized such that $C^{(s)}(0)=1$)
for $L=80$, $K=2$ and $K'=0 \rm ~and~ 0.02$ 
obtained by QMC. A clear enhancement of the AF 
correlations can be seen when a soliton-impurity bound
state is present ($K'= 0.02$) in contrast to the case of
an almost uniform pattern with strong end bonds shown in 
Fig. \ref{disteven} ($K'=0.0$). This result is
consistent with the enhancement of AF correlations close to an
impurity for weak end bonds found in Ref. \onlinecite{Laukamp}.
One expects that this increase of the AF correlations in the vicinity 
of impurities carrying an effective spin-1/2 magnetic moment can be 
observed in inelastic neutron experiments at low energy transfer and
momentum transfer $q\sim \pi$. 

Our results are also consistent with the experimental observation
of Curie laws in the susceptibility of Zn-doped 
(spin-0)~\cite{matsuda} and Ni-doped (spin-1)~\cite{koide} CuGeO$_3$
materials. 
Note however that, in the case of Zn-doping, due to the mechanism
described in Sec.~\ref{inter_cou}, a significant fraction of the 
impurities may not carry an effective spin-1/2 leading to a 
reduction of the magnitude of the Curie

\noindent
term. In the case of Ni-doped
CuGeO$_3$~(Ref.~\onlinecite{grenier}), an interesting cross-over at
low temperature was observed between two different Curie law
behaviors. We interpret this low temperature behavior with the
formation of an effective singlet at the Ni-site which can bind a
spin-$\frac{1}{2}$ soliton as described in Sec.~\ref{1d}.

\section{Conclusion}
In this work, we have studied several spin-lattice models using various
numerical techniques. Local non-uniform dimerization
patterns have been obtained resulting from inhomogeneities due 
to impurity doping. We have shown that the lattice responds strongly to
the variation of the spin density even in the adiabatic case.
Binding between elementary solitonic
excitations and spin-S impurity (S=0 or 1) has been
investigated. In strictly 1D models, 
no binding between solitons and non-magnetic impurities occurs.
However, a small binding between spin-1 impurities and solitons has been
inferred. 
The role of the three-dimensionality of the lattice was also investigated
by comparing a model with a fixed mean field-like dimerization to a model with 
an explicit coupling to an adiabatic lattice which can locally
relax to follow the magnetic order. 
In the latter case, the interchain elastic coupling was shown to be 
responsible for the binding of solitons to spin-0 impurities.
However, in contrast to the dimerized Heisenberg model, we predict
in this model the existence of $S=0$ impurities carrying no spin-1/2
solitonic excitation. These impurity sites must appear 
in pairs separated by sufficiently
small chain segments with an even number of sites and a dimerization pattern
in antiphase with respect to the 3D dimerization order.

\section{acknowledgements}

One of us (J.~R.) wishes to thanks A. Dobry for many useful discussions.
We thank IDRIS (Orsay, France) for allocation of CPU time on the C94 and C98
CRAY supercomputers.
The use of the computing facilities at the Supercomputer Computations
Research Institute, Tallahassee, Florida, is also gratefully
acknowledged.

\end{document}